\documentclass[prl,twocolumn,amsmath,amssymb,showpacs,superscriptaddress]{revtex4}

\newcommand{\csixty}{C$_{60}$}

\newcommand{\csixtyfourminus}{C$_{60}^{4-}$}
\newcommand{\csixtysixminus}{C$_{60}^{6-}$}
\newcommand{\kthreecsixty}{K$_3$C$_{60}$}
\newcommand{\kfourcsixty}{K$_4$C$_{60}$}
\newcommand{\ksixcsixty}{K$_6$C$_{60}$}
\newcommand{\kncsixty}{K$_n$C$_{60}$}
\newcommand{\kplus}{K$^+$}
\newcommand{\eq}[1]{~(\ref{#1})}

\newcommand{\tu}[1]{^{\textrm{#1}}}
\newcommand{\tl}[1]{_{\textrm{#1}}}
\newcommand{\V}[1]{{\bf #1}}
\newcommand{\ket}[1]{|#1\rangle}
\newcommand{\bra}[1]{\langle#1|}

\usepackage{graphics}
\usepackage{bbm}
\usepackage{psfig}
\usepackage{epsfig}
\usepackage{dcolumn}
\usepackage{bm}

\begin{document}
\title{Theory of the different photoemission spectra of metallic and
  insulating \csixty{} compounds}
\author{Samuel Wehrli}
\email{swehrli@phys.ethz.ch}
\author{T.M.~Rice}
\author{Manfred Sigrist}

\affiliation{Theoretische Physik, ETH-H\"onggerberg, CH-8093 Z\"urich, 
             Switzerland}
\date{\today}

\begin{abstract}
Metallic K$_3$C$_{60}$ shows pronounced structure and a sharp
Fermi edge in integrated photoemission spectra (PES) while the insulating 
K$_4$C$_{60}$ and K$_6$C$_{60}$ phases display 
only a broad structureless peak. We find that both types of 
spectra can be explained by the 
coupling to the optic
vibrations  of the K$^+$/C$_{60}^{n-}$ ionic lattice. This is suppressed in
K$_3$C$_{60}$ 
due to metallic screening but is strong in the insulating 
phases. We use the
non-crossing approximation to calculate the density of
states (DOS) of electrons in K$_3$C$_{60}$ 
coupled to the intramolecular $H_g$ modes 
in good agreement with the experiment. For K$_4$C$_{60}$ and K$_6$C$_{60}$ 
strong coupling to the low energy optic K$^+$/C$_{60}^{n-}$ modes controls 
the DOS and yields broad peaks in the PES. A moment expansion is used to
calculate the  position and width of these peaks which
agree well with the experiment.
\end{abstract}


\maketitle


Alkali-metal-doped \csixty{} has been investigated extensively by
photoemission spectroscopy (PES) in bulk systems
(Ref.~\onlinecite{chen91,benning93,hesper00} and references therein). Recently,
Yang et al. measured the electronic band dispersion of
\kthreecsixty{} in a monolayer system on Ag(111) by angle resolved
PES (ARPES)~\cite{shen03}.    
By comparing angle integrated spectra (AIPES) from several experiments a
generic feature emerges which is the striking difference between the spectra of metallic
\kthreecsixty{} and the insulating \kfourcsixty{} and \ksixcsixty{} (the
former is a Jahn-Teller distorted insulator, the latter  a band
insulator): \kthreecsixty{} shows a metallic spectra with a
sharp Fermi edge and distinct structures
whereas the insulating phases display a broad Gaussian-peak. In the
present work we address this difference. The key ingredient is the
large change in coupling strength 
of the low-energy ($\omega\approx 10$~meV) optic vibrations
of the \kplus{} ions. These modes cause a net shift of the \csixty{} 
molecular levels and their interaction is
efficiently screened in metallic \kthreecsixty{} 
where it is reduced by a factor
$10^{-4}-10^{-2}$~\cite{gunnarsson92,koch99}. However, in the insulating
phases, where metallic screening is absent, they couple strongly ($g\approx
70$~meV, see below) due to the direct Coulomb interaction of the ionic
charge with the photoemission hole. The resulting physics is best illustrated
by the toy-model 
$H=\omega\, a^\dagger a+g\,c^\dagger c\,(a^\dagger+a)$ where a single electron
is coupled to a harmonic oscillator. The corresponding photoemission spectrum
is a Poisson distribution 
$P(\epsilon)=\sum_n \nu^ne^{-\nu}/n!\,
\delta(\epsilon\!+\![n\!-\!\nu]\omega)$ 
where
$\nu=(g/\omega)^2$ is the average number of
excited phonons emitted 
during the photoemission process. In the insulating phases
$\nu$ is very large ($\nu\approx 50$) 
and as a result the spectrum becomes incoherent 
resulting in a Gaussian-like shape. In the following we treat the metallic and
insulating phases separately. First, using the non-crossing
approximation (NCA), it is shown that the spectrum of metallic
\kthreecsixty{} is dominated by the coupling to the intramolecular \csixty{}
modes. Second, we use moment expansion to calculate position and width of the
Gaussian-like spectra of \kfourcsixty{} and \ksixcsixty{}.
 
\begin{figure}
  \begin{center} 
  \includegraphics[width=0.9\columnwidth]{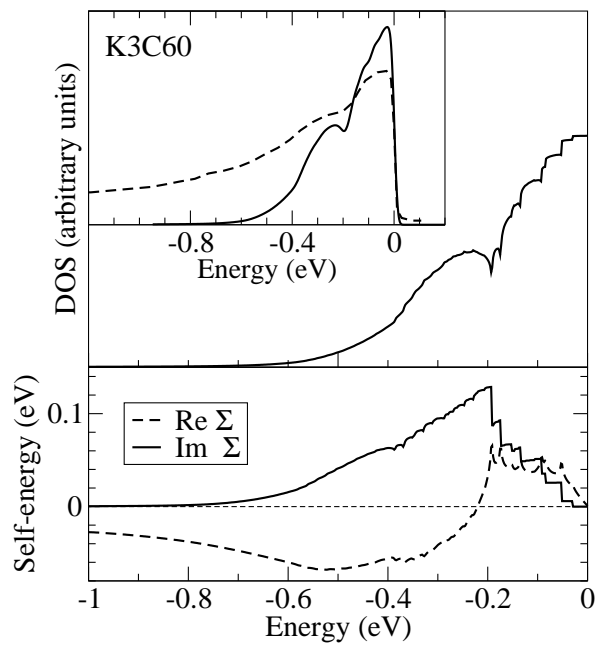}
  \end{center}\vspace{-0.7cm}

  \caption{\label{fig:c3}
    Solution of the NCA at half filling ($\mu=0$) which corresponds to
    \kthreecsixty{}.
    \emph{Upper panel:} Occupied part of the interacting DOS.
    \emph{Lower panel:} Advanced self-energy. 
    \emph{Inset:} Occupied DOS (solid line) 
    convoluted with a Gaussian ($\sigma=10$~meV) and compared to the 
    experimental spectrum
    of the monolayer (dashed line)~\cite{shen03}. }
\end{figure}

The only modes that couple strongly ($\lambda\approx 1$) in \kthreecsixty{}
are the intramolecular $H_g$ modes which cause a splitting of the threefold
degenerate LUMO's of \csixty{} (See Ref.~\onlinecite{gunnarsson97} for more
details). We treat these modes by NCA~\cite{engelsberg63} and neglect the
on-site Coulomb interaction ($U\approx 1$~eV in bulk). This is justified in
a monolayer adsorbed on Ag(111) where $U$ is reduced by the proximity of the
metal~\cite{hesper97}. Here we restrict our attention to such systems.
A similar calculation was performed by Liechtenstein et al. who showed
that the width of the plasmon in \kthreecsixty{} can be explained by 
electron-phonon coupling~\cite{liechtenstein96}. 
The Hamiltonian which 
describes the conduction band of \kthreecsixty{} coupled to the $H_g$ modes is
(setting $\hbar=1$)
\begin{eqnarray} \label{H}
  H_{H_g-\textrm{vib} } &=& \sum_{j\delta\; nm} t_{nm}(\delta)\,
    c^\dagger_{j\!+\!\delta m}c_{jn} +
    \sum_{j\nu k} \omega_{\nu}\,a^\dagger_{j\nu k}a_{j\nu k} +     
    \nonumber \\
    & &\sum_{jnm\atop \nu k}g_{\nu}\,
        c^\dagger_{jm}c_{jn}
        \left[C_{nm}^k\,a_{j\nu k}^\dagger+
             C_{mn}^k\, a_{j \nu k}\right].
\end{eqnarray}
The first term is the tight-binding band in standard notation formed by the  
3-fold degenerate ($t_{1u}$) LUMO's of \csixty{}.
The sum is over the orbitals $n$ and $m$, the lattice sites $j$
and the nearest neighbors $\delta$.   
As the spin orientation is preserved in the Hamiltonian,
explicit sums over spins are
dropped throughout.
The second term is the energy $\omega_\nu$ of the 
eight five-fold degenerate $H_g$ vibrational multiplets~\cite{gunnarsson97}. 
The indices $\nu$ and  $k$ denote the multiplet and the mode
respectively. The last term describes the electron-phonon coupling.
The phonon energies $\omega_\nu$ and the coupling parameters $g_\nu$
are dispersionless.
The parameters $g_\nu$ are related to the partial 
coupling constant $\lambda_\nu$ by  
$g_\nu^2 = \frac{3}{2}\omega_\nu\lambda_\nu/N(0)$
where $N(0)$ is the density of states
per spin and molecule~\cite{lannoo91}.
Values for $\omega_\nu$ and 
$\lambda_\nu$ where taken from Ref.~\onlinecite{gunnarsson95}.
The structure of the
coupling is given by the coefficients 
$C_{nm}^k = \sqrt{\frac{3}{5}}\,(-1)^m \langle 2,k | 1,-m;1,n \rangle$
where $\langle \ldots \rangle$ is the Clebsch-Gordan
coefficient~\cite{tosatti94}. 
The normalization is such that $\sum_{knm}\left(C_{nm}^k\right)^2=3$.
In the NCA, the electron self-energy 
is determined self-consistently by the lowest order self-energy diagram with
the interacting Green's function. In the present
problem the Green's function $G_{nm}(\omega,\V k)$ and self-energy
$\Sigma_{nm}(\omega,\V k)$
are $3\times3$ matrices. The free phonon propagator
$D^0_{\nu k}(\omega, \V q)=D^0_\nu(\omega)$ depends only on
frequency and the multiplet. As a consequence, only the local part of the
Green's function $G\tu{loc}_{nm}=1/N\sum_{\V k}G_{nm}(\omega,\V k)$ enters
which renders the
self-energy 
$\Sigma_{nm}(\omega,\V k)=\Sigma_{nm}(\omega)$ local as well. This
is a consequence of including only non-crossing diagrams. 
The evaluation of the basic diagram can be simplified further when the symmetry
of the lattice is
used. The momentum-independent
Green's function $G_{nm}\tu{loc}$ and   
self-energy $\Sigma_{nm}\tu{loc}$ 
have to be invariant under all symmetry transformations
which belong to the point-group of the lattice. In particular, 
in a cubic environment (such as a fcc lattice)  
\begin{eqnarray}\label{nca6}
   G_{nm}\tu{loc}(\omega)&=&\hat G\tu{loc}(\omega)\,\delta_{nm},\quad 
  \Sigma_{nm}(\omega)=\hat \Sigma(\omega)\,\delta_{nm},
\end{eqnarray}
where
$\hat G\tu{loc}(\omega)$ and 
$\hat \Sigma(\omega)$ are scalars. 
On the surface  or in a monolayer the symmetry is lower than cubic  
and the Green's function
has additional off-diagonal parts.
However, the corresponding corrections were found to be 
small (1~\% or less) and 
therefore it is an excellent approximation to 
use the Green's function and self-energy as
given in\eq{nca6}. This yields the scalar
equation 
\begin{eqnarray}\label{nca8}
  \hat \Sigma(E) &=& i\sum_{\nu} g_\nu^2 \int \frac{d\omega}{2\pi}\,
   D_0^\nu(\omega)\, \hat G\tu{loc}(E-\omega).  
\end{eqnarray}
Thus, the problem is simplified to a single
band interacting with a discrete set of phonon modes. Relation\eq{nca8},
which is an equation for the self-energy,
is solved iteratively  using the advanced Green's function. 
In\eq{nca8}, the band-structure only enters via the density of states (DOS) 
and 
we chose a generic square DOS with a width $W=0.5$~eV. 
Using other bare DOS revealed
that the interacting DOS depends only weakly on the form of the  bare DOS.  
The result for half-filling ($\mu=0$), which corresponds to
\kthreecsixty{}, is shown in Fig.~\ref{fig:c3}. 
The interacting DOS shows an overall structure, such as a dip at 0.2~eV and a
second hump at 0.4~eV, which agrees well with the AIPES of the monolayer 
system~\cite{shen03}. 
\begin{table}
\begin{tabular}{l|c|c|c}
  Mode & $\omega_\nu$ & $\lambda_\nu/N(0)$  
     & $g _\nu$ \\
  \hline
  $H_g(8)$ & 195 &  23 & 82 \\
  $H_g(7)$ & 177 &  17 & 67 \\
  $H_g(6)$ & 155 &   5 & 34 \\
  $H_g(5)$ & 136 &  12 & 50 \\
  $H_g(4)$ &  96 &  18 & 51 \\ 
  $H_g(3)$ &  88 &  13 & 41 \\
  $H_g(2)$ &  54 &  40 & 57 \\
  $H_g(1)$ &  34 &  19 & 31 \\
  \hline
  $A_g(2)$ & 182 &  11 & 55 \\
  $A_g(1)$ &  61 &   0 &  0 \\
  \hline
  K-mode  &  8.9$\tu a$, 10.9$\tu b$ &   - & 65$\tu a$, 72$\tu b$ \\
  Acoustic & 3.8 & - & 10
 \end{tabular}
\caption{\label{tab:phpar}
  Frequencies and coupling constants for the vibrational modes in
  \kncsixty{} (all energies are in meV).
  Parameters for the intramolecular $H_g$ and $A_g$ modes were 
  taken from Ref.~\onlinecite{gunnarsson95}. 
  The coupling constant for the lattice vibrations
  were calculated in this work.
  In \kthreecsixty{} the coupling to the $A_g$ modes
  and the vibrations of the
  ionic lattice  is efficiently
  suppressed by metallic screening. 
  a) applies for \kfourcsixty.  b) applies for \ksixcsixty.
   }
\end{table}
\begin{figure}
  \begin{center} 
  \includegraphics[width=0.8\columnwidth]{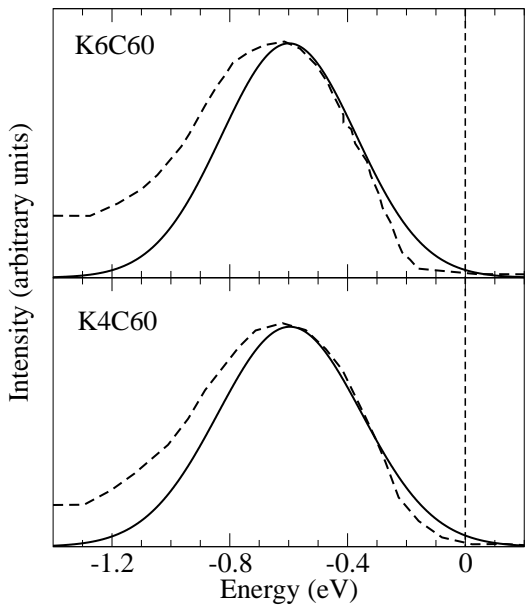}
  \end{center}\vspace{-0.7cm}

  \caption{\label{fig:pol}
    Gaussian fit (solid curves) to the photoemission spectrum using the first
    and second moment as calculated in\eq{mom} and plotted with respect to the
    chemical potential. The dashed curves are the experimental data
    (Ref.~\onlinecite{benning93} for \ksixcsixty,
    Ref.~\onlinecite{hesper00} for \kfourcsixty) }
\end{figure}
%


The insulating compounds \kfourcsixty{} and  \ksixcsixty{} differ importantly
from \kthreecsixty{}. As discussed in the introduction, in the absence of
metallic screening, the coupling to the \kplus{} modes is strong
and the ground
state of a hole created by photoemission will be polaronic. This type
of physics is not captured by NCA. 
On the other hand, the ground states of the insulating phases are
rather simple: \ksixcsixty{} is a trivial band insulator, whereas in 
\kfourcsixty{} there are four localized electrons on each \csixty{} which form
a singlet due to Jahn-Teller distortions~\cite{tosatti94,gunnarsson95b}. 
If the ground state $\ket{\Psi_0}$ is known, the moments 
$\mu_k = \int d\epsilon \,\epsilon^k\, P(\epsilon)$ of the spectrum 
$P(\epsilon)$ (normalized to 1) can exactly be calculated by evaluating the
expectation values 
\begin{eqnarray}\label{mom} 
  \mu_k  
   =\frac{1}{N}\sum_{jn} \bra{\Psi_0} 
     c_{jn}^\dagger[[c_{jn},\underbrace{H],H\ldots]}_k\ket{\Psi_0}. 
\end{eqnarray} 
The sum is over 
all sites $j$ and orbitals $n$ and $N$  is
the total number of the electrons. $H$ is the full Hamiltonian and includes
all vibrational modes as well as Coulomb interactions. 
Although relation\eq{mom} is exact, the reconstruction of a distribution
from a finite number of known moments is an ill-defined problem if the overall
shape of the distribution is unknown. However, as was argued above, the strong
coupling to the low-energy optic modes yields a very incoherent, and therefore
Gaussian-like spectrum. Hence, for this physical reason the
distribution should be well approximated by a 
Gaussian which is determined by the first and second moment as given
by\eq{mom}. The procedure which follows is, first, to
determine all contributions to $H$ and, second, calculate the moments
by\eq{mom}.  

The additional phonon contributions to $H$ are 
modes which cause a net shift of
the molecular orbitals and 
therefore are no longer screened in the insulating
phases~\cite{gunnarsson92,koch99}. These modes are 
the intramolecular $A_g$ modes and the vibrations of the
ionic lattice. 
In principle, lattice vibrations also couple via a change in hopping, but
the corresponding coupling is much smaller~\cite{gunnarsson97}. 
In the case of the two
intramolecular $A_g$, 
the frequencies $\omega_\nu$ and coupling parameters $g_\nu$ 
are dispersionless. As for the $H_g$ modes, the latter is given by
$g_\nu^2 = \frac{3}{2}\omega_\nu\lambda_\nu/N(0)$ (see above) and 
values were taken from Ref.~\onlinecite{gunnarsson95}.
The coupling to the ionic lattice has been
studied much less extensively, mainly because it is negligible in the
superconducting \kthreecsixty{}. Here we consider the coupling due
to the Coulomb interaction between the ionic charges. 
The mass ratio 
$M_{\textrm{\csixty}}/M_{\textrm{K}}=18.4$ is big which allows to separate the
lattice vibrations into optic dispersionless $K$-modes and acoustic modes.
The frequencies of these modes were measured by EELS~\cite{silien03} where it
was observed that \kplus{} ions close to the surface have substantially
lower frequencies. 
As photoemission is surface sensitive, we use these values which
are  
$\omega_K=8.9$~meV and $\omega_K=10.9$~meV for \kfourcsixty{} and
\ksixcsixty{} respectively.   
In what follows, only the averaged coupling constant
$\bar g_K^2=N_s^{-1}\sum_{\V q\alpha} |g_{\V q\alpha}|^2$ 
enters (the sum is over
all optic K-modes $\alpha$) which is given by (with $\hbar$ exceptionally included) 
\begin{equation}\label{gk}
  \bar g_K^2 =
  \frac{e^2\hbar}{2 M_K\omega_K}\sum_{r\in\{\textrm{K}^+\}}\V E_{r}^2.
\end{equation}
The sum
runs over all \kplus{} ions $r$ and $\V E_r$ is the electric field at the
positions $r$ and caused by an additional hole on the \csixty{} 
molecule at the origin. The sum in\eq{gk} depends on the lattice.  
\ksixcsixty{} has a bcc lattice with a cubic lattice constant
$a=11.39$~\AA{} where 
each \csixty{} molecule is surrounded by 24
\kplus{} ions located at $(0,0.5,0.25)a$~\cite{erwin91}. 
Considering  the bare coupling of the
closest by \kplus{} ions yields $g_K=120$~meV. 
Taking into account the polarizability of 
the \csixtysixminus{} ions by
multipole expansion~\cite{wehrli03} reduces the
electric fields entering\eq{gk} by 40~\% and yields a 
coupling constant $g_K=72$~meV.
In \kfourcsixty{} distances between \csixtyfourminus{} and \kplus{} are almost
the same as in \ksixcsixty{}, however,  
every \csixtyfourminus{} ion is surrounded by 16 \kplus{} ions instead of 24.
This reduces $g_K$ in \kfourcsixty{} by a factor $\sqrt{2/3}$ with respect
to \ksixcsixty{}. Taking also into account the difference in $\omega_K$ we
find $g_K=65$~meV for \kfourcsixty{}.
Coupling to the acoustic modes is 
more involved because both  
$\omega_{\V q\alpha}$ and $g_{\V q\alpha}$
are $\V q$ dependent. We used a simple spring model 
parametrized by a phonon frequency of 5~meV at the Brillouin zone
boundary~\cite{pintschovius96}. This yields an 
averaged coupling constant $\bar g_a=10$~meV. In addition we find an average
frequency $\bar\omega_a= N_s^{-1}\sum_{\V q\alpha}\omega_{\V q\alpha} 
|g_{\V q\alpha}/\bar g_a|^2 =3.8$~meV. Finally, the Coulomb interactions
should also be included in $H$. However, they vanish in the case of
\ksixcsixty{} where a single hole is created in a full band. In \kfourcsixty{}
there is a 
contribution from the on-site Hunds-rule coupling term which involves
an exchange energy $K=50$~meV (see Ref.~\onlinecite{capone00} for a
detailed description). This term leads to an additional increase of the width
of the \kfourcsixty{} spectrum.

In the following we will discuss the results for \ksixcsixty{} in more detail
than those of \kfourcsixty{} because the ground state of the latter involves
Jahn-Teller distorted molecules and a detailed discussion would go beyond the
scope of this letter.  As mentioned above, 
the ground state of \ksixcsixty{} is a full band and
trivially given by
$\ket{\Psi_0}=\prod_{jn}c^\dagger_{jn}\ket{\textrm{vac}}$. 
Note that $\ket{\Psi_0}$ doesn't have any phonon excitations.
Coulomb terms can be neglected for \ksixcsixty{} and the Hamiltonian 
$H=H\tl{kin}+H_p+H_{ep}$ consists of the kinetic energy, the phonon energies
and the electron-phonon coupling terms. As in the previous section we assume a
quadratic bare DOS $\rho_0$ with a width $W=0.5$~eV and centered
around zero. Using relation\eq{mom} to evaluate the moments of the
photoemission spectrum one finds that $\mu_1=0$. The second moment is 
$\mu_2 = \mu_2(\rho_0)+\sum_x g_x^2$ where $\mu_2(\rho_0)=W/(2\sqrt 3)$ is the
second moments of the normalized square DOS $\rho_0$. The sum is
over all phonon modes $x$. Similarly, the third moment is 
$\mu_3 = -\sum_x w_x g_x$. Using the parameters as listed in 
Tab.~\ref{tab:phpar} one finds $\sigma=\sqrt{\mu_2}=0.229$~eV and 
$\mu_3/\sigma^3=-0.319$. Usually, the photoemission spectrum 
is plotted with respect to the
chemical potential which, per definition, is 
$\mu=E_{0}(N)-E_0(N-1)=-E_0(N-1)$ and therefore corresponds
to the polaron ground state energy. In the small polaron limit, where the hole
is localized on a single molecule, the chemical potential is given by the
relaxation energy of the phonon degrees of freedom which couple to the hole: 
\begin{equation}\label{mu}
 \mu = -E_0(\textrm{C}_{60}^{1-}) + 
       \frac{\bar g_K^2}{\omega_K} +\frac{\bar g_a^2}{\bar \omega_a}
     = 0.599\textrm{~eV},
\end{equation}
$E_0(\textrm{C}_{60}^{1-})$ is the
Jahn-Teller ground
state energy of an electron in the LUMO of an isolated \csixty{} molecule
interacting with the intramolecular modes.
By particle-hole symmetry, this 
is the same as $E_0(\textrm{C}_{60}^{5-})$ and
was calculated numerically in Ref.~\onlinecite{gunnarsson95b}. The last two
terms are the energy gain due to the ionic lattice distortions. It must be
noted that the total
contribution of the K-modes to the chemical potential 
is proportional to $w_K^{-2}$,
because $g_K^2$ itself is proportional to $w_K^{-1}$ (relation\eq{gk}). Hence,
$\mu$ depends sensitively on $\omega_K$.
Note that this estimate is a lower bound for the chemical potential, because
in the insulating \ksixcsixty{} 
it may lay everywhere in the bandgap rather than on
the top of the filled band. In Fig.~\ref{fig:pol} a Gaussian of width $\sigma$
and shifted by $\mu$ is plotted. The result compares well with the experimental
curve from bulk measurements although the width is somewhat too
small~\cite{benning93}. 
Other fitting functions which accounted correctly for the non-zero
third moment were tested as well which, however, yielded only slightly
different curves than the Gaussian shown here. 

The case of \kfourcsixty{} is more complicated, due to Jahn-Teller distorted
molecules where the three-fold degenerate LUMO is split into a singlet and a
doublet. In \kfourcsixty{} the doublet is lower in energy and occupied by 
four electrons. Hence, they form a full band in the solid which results in the
insulating state. The problem of the Jahn-Teller distortion in \csixty{}, which
cannot be solved analytically, has
been extensively studied by various
approaches~\cite{tosatti94,gunnarsson95b}. In addition, we developed
a variational wavefunction for \csixtyfourminus{} which 
yields a ground state energy in agreement with exact diagonalization
results~\cite{gunnarsson95b}. In order to calculate the expectation values
in\eq{mom} we use this variational ground state
(details of the calculation of the will be published elsewhere).
We find a first moment $\mu_1=-0.174$~eV which is due to  
the energy gain of the
Jahn-Teller distortion. The chemical potential is  again 
given by\eq{mu}, except that
$-E_0(\textrm{C}_{60}^{1-})$ has to be replaced by 
$E_0(\textrm{C}_{60}^{4-})-E_0(\textrm{C}_{60}^{3-})$
which yields $\mu=0.422$~eV~\cite{gunnarsson95b}. Finally, the second moment
is $\sigma=\sqrt{\mu_2}=0.244$~eV which is almost the same as in
\ksixcsixty{}. In Fig.~\ref{fig:pol} the corresponding Gaussian is plotted and
compared to the experimental 
data from bulk measurements~\cite{hesper00}. Again,
good agreement is found.

In conclusion, we showed that the different 
photoemission spectra in metallic and
insulating \kncsixty{} are due to a large change in the  
coupling strength to the low-energy, optic
\kplus{} modes. Theoretical calculations for both cases yield good results.

We are grateful to E. Koch and O. Gunnarsson for useful
discussions and, especially, to  Z.-X.~Shen for drawing our
attention to this problem as well as to W.L. Yang for providing the
experimental data.
We also acknowledge the support from the Swiss Nationalfonds. 

\end{document}